# A Hypothesis on Good Practices for AI-based Systems for Financial Time Series Forecasting: Towards Domain-Driven XAI Methods

Branka Hadji Misheva and Joerg Osterrieder

**Abstract**

Machine learning and deep learning have become increasingly prevalent in financial prediction and forecasting tasks, offering advantages such as enhanced customer experience, democratizing financial services, improving consumer protection, and enhancing risk management. However, these complex models often lack transparency and interpretability, making them challenging to use in sensitive domains like finance. This has led to the rise of eXplainable Artificial Intelligence (XAI) methods aimed at creating models that are easily understood by humans. Classical XAI methods, such as LIME and SHAP, have been developed to provide explanations for complex models. While these methods have made significant contributions, they also have limitations, including computational complexity, inherent model bias, sensitivity to data sampling, and challenges in dealing with feature dependence. In this context, this paper explores good practices for deploying explainability in AI-based systems for finance, emphasizing the importance of data quality, audience-specific methods, consideration of data properties, and the stability of explanations. These practices aim to address the unique challenges and requirements of the financial industry and guide the development of effective XAI tools.

## 1. Introduction

Machine learning (ML) and deep learning (DL) have gained significant popularity in various aspects of data science and has increasingly found its place in prediction tasks for financial and economic problem sets (Dingli and Fournier, 2017, Persio and Honchar, 2017, Sen and Mehtab, 2021, Yang, 2021). Many empirical studies provide evidence of the many advantages state-of-art AI-based technologies can bring to the financial sector including enhancing customer experience (Bhattacharya and Sinham, 2022), democratizing financial services (Athey, 2019, Bazarbash, 2019, Sadok et al., 2022), improving consumer protection (Alzahrani and Aljabri, 2022, Aslam et al., 2022) and significantly improving risk management (Guidici, Hadji Misheva and Spelta, 2020, Ahelegbey, Guidici and Hadji Misheva, 2019, Danielsson et al., 2022). ML and DL methods have also been applied to stock and forex markets, and (in many cases) have demonstrated substantial superiority over traditional approaches (Jung and Choi, 2021, Yildirum et al., 2021, Korczak and Hernes, 2017, Hu et al., 2021). This conclusion has been reinforced also by recent occurrences in the Makridakis Forecasting Competitions (M4 and M5), where a combination of Exponential Smoothing Recurrent Neural Network and LightGBM claimed victory in their respective competitions (Makridakis 2020 and Makridakis 2021).

While the integration of ML and DL techniques into financial prediction and forecasting tasks holds the promise of enhanced predictive accuracy, this advantage comes at the expense of increased complexity and reduced interpretability. Such complex methods are often characterized as 'black-boxes' because comprehending how variables interact to produce a specific outcome can oftentimes be very challenging. The limited transparency impacts the reliability of ML and DL models and the willingness of industry experts to employ them in sensitive domains like finance. Consequently, there has been a substantial surge of interest in the field of **eXplainable Artificial Intelligence (XAI)** over the past few years (Linardatos et al., 2021, Dazeley et al., 2021, Gunning et al., 2019). The main objective of XAI is to create models that can be easily understood by humans, a property that is particularly critical in fields like finance and healthcare. In these sectors, experts require assistance in solving problems more efficiently, and they also seek meaningful results that they can comprehend and have confidence in. This growing interest has led to the development of various

XAI methods. Every year, a large number of studies are published that introduce processes and guidelines that would allow users to comprehend and trust the outputs created by ML and DL algorithms (Montavon et al., 2018, Mittelstadt, 2019, O'Sullivan, 2019, Pearl, 2019). At the same time, various studies have attempted to operationalize those guidelines and propose methods that can bring explanations to users (Ribeiro et al., 2016, Montavon et al., 2017, Ribeiro et al., 2018, Lundberg and Lee., 2017, Lundberg et al., 2018 and 2020, Kauffmann et al., 2019, Luo, 2020, Montavon et al., 2022, Kole et al., 2022). In general, these methods are classified in view of four main criteria (Linardatos et al., 2021): (i) the type of algorithm on which they can be applied (where we distinguish between model-specific and model-agnostic XAI methods); (ii) the unit being explained (if the method provides an explanation which is instance-specific then this is a local explainability technique and if the method attempts to explains the behavior of the entire model, then this is a global explainability technique), (iii) the data types (tabular vs text vs images-specific XAI methods), and (iv) the purpose of explainability (ex. XAI methods that test sensitivities or improve model performance, etc.). Some of these methods are discussed in Section 2.

Although many emerging XAI tools have an established utility in brining explainability to AI-based systems build in certain use cases, their applicability to complex models developed for financial data in general, and financial time series in particular, remains rather limited. A survey conducted by Weber at al. (2023) finds that only 16 papers were published in 2021 that deal with XAI applications in finance and those include diverse thematic focus, application, and methodologies (qualitative reviews, experiments, case studies, empirical work). The study further reveals an imbalance in the disciplines from which the research articles originate, with a majority originating from information systems and computer science, and only a quarter from the finance literature. This suggests a need for greater engagement of the finance discipline in XAI research, given the specific requirements of the industry. Additionally, the study finds that there is a significant number of articles that are published in outlets with no ratings, possibly indicating a focus on practical applications over theory-building and development.

In this context, the purpose of this work is to provide a critical viewpoint on the utility of classical, state-of-art methods for AI-based systems applied to finance, and thus deliver theoretical and practically relevant guidelines on how XAI tools for finance should be build.

## 2. Utility of Classical XAI methods for AI-based systems applied to financial problem sets

### *2.1. Overview of Classical XAI methods*

In this section, we aim to provide an overview of the clssicial, state-of-art XAI methods available in the literature. To accomplish this goal, we reviewed XAI papers from the last three years (2019–2021) referencing, containing, reviewing or proposing taxonomies of explainability methods (among which: Arrieta et al., 2019, Gunning et al., 2019, Linardatos et al., 2021, Dazeley et al., 2021 and Saeed and Omlin, 2023). While this is by no means a systematic review, we focused on representative papers in the field and as such we consider the points we make to be sufficiently general to be applicable to taxonomies that have not been considered.

To start, the literature offers a clear distinction between directly intrinsically explainable models and the need for explaining a model after it has been trained. The former is sometimes called **transparent model by design**. A clear example of an intrinsically explainable model is the linear regression. By design, the linear regression provides estimates or coefficients that provide valuable information about the relationship between the independent variables and the dependent variable. Specifically, these estimates tell us both the magnitude (the size of the effect) and direction (positive or negative) of the impact each independent variable has on the dependent variable included in the model specification. Such estimates make it easy to explain how changes in input variables affect the output. Even in situations where we need to model non-linear relationships, it is possible to train models that are transparent. One such example is the use of decision trees. Decision trees are a type of model that can capture complex, non-linear relationships in the data while remaining transparent and interpretable. Namely, decision trees are explainable because they provide a clear and interpretable representation of decision-making processes by breaking down complex decisions into a sequence of easily understandable binary choices or conditions. This transparency allows users to trace the logic behind each decision path and gain insights into how specific features or variables influence the final outcome.

A key question then becomes, when do we run into an explainability constraint? We run into explainability constraints primarily when dealing with high-dimensional data, where the sheer number of features or variables makes it challenging to comprehend the decision-making process. In such cases, the complexity of the model may hinder our ability to provide clear and intuitive explanations for the relationships between input features and the model's predictions, potentially reducing our understanding of the underlying patterns in the data. In such cases, when models fail to satisfy the established criteria for transparency, an alternative method must be formulated and applied to uncover the rationale behind the model's decisions. This objective is achieved through the utilization of post-hoc explainability techniques, also known as **post-modeling explainability**. These techniques aim to convey intelligible insights into how a pre-existing (complex ML or DL) model generates predictions for a given input. As stated previously, there are different taxonomies of XAI methods. The literature primarily distinguishes post-hoc explainability techniques based on model that they could be applied to. In this context, XAI methods are grouped into: (i) approaches designed for general application to a wide range of ML or DL models, i.e., model agnostic post-hoc techniques, and (ii) approaches specifically tailored for a particular ML or DL model, and therefore, not readily applicable to other types of learners, i.e., model-specific post-hoc explainability techniques. A further distinction between XAI techniques is based on the unit for which the explanation is provided. Global post-hoc explainability techniques aim at providing insight into the overall rational of the model whereas local post-hoc methods focus on explaining the model's behavior at the individual unit of analysis. Among the global, model-agnostic methods, we find techniques like feature importance plots, partial dependency plots (PDP), accumulated local effects plots, and global surrogate models. Feature importance plots remain one of the most widely used method for understanding the innerworking of complex ML and DL models. These plots help users understand which features have the most substantial influence on the model's outcomes and can be crucial for tasks like feature selection, model debugging, and gaining a better understanding of the underlying data relationships. Although feature importance plots can be very useful for identifying which inputs have a significant impact on the model's performance, their utility is limited as they do not tell us anything about the relationship that emerges between the features and the target. This limitation is further addressed by techniques such as the partial dependency plots (PDPs). PDPs are visualization techniques used to illustrate the relationship between a specific input feature and the predicted outcome of a model while keeping all other features constant or at fixed values. PDPs provide valuable insights into how a single feature affects the model's predictions, allowing for a more in-depth understanding of its impact on the model's behavior. As indicated, they address a fundamental limitation of feature importance plots by providing a detailed representation of the relationship between a specific input feature and the model's predictions while holding all other features constant. This approach mitigates the potential shortcomings of feature importance plots, such as the inability to capture complex feature interactions and dependencies, thus offering a more comprehensive and nuanced understanding of how individual features influence the model's behavior. Yet, a key challenge of the PDPs is their underlining assumption of independence. Namely, PDPs assume that the feature(s) for which the partial dependence is computed are not correlated with other features and this is oftentimes not the case in real-world applications.

Among the model agnostic XAI techniques, two frameworks have been widely recognized as the state-of-the-art methods for obtaining explanations (Krishna et al. 2022) and those are: (i) the **LIME framework**, introduced by Ribeiro et al. in 2016, and (ii) **SHAP values**, introduced by Lundberg and Lee in 2017. LIME, short for locally interpretable model agnostic explanations, is an explanation technique which aims to identify an interpretable model over the representation of the data that is locally faithful to the classifier (Ribeiro et al. 2016). Specifically, LIME disregards the global view of the dependence between the input-output pairs and instead derives a local, interpretable model using sample data points that are in proximity of the instance to be explained. The specific steps in obtaining the explanations are:

1. **Instance selection**: LIME initiates by selecting a specific data instance of interest for which an explanation is desired. This instance serves as the focal point for the interpretability analysis.
2. **Perturbed data generation**: To gain insight into the model's behavior in the vicinity of the chosen instance, LIME generates a dataset comprising perturbed versions of the original instance. These perturbations typically involve introducing slight variations in feature values while preserving the essential characteristics of the selected instance.
3. **Prediction acquisition**: For each of the perturbed instances, LIME solicits predictions from the target black-box model. The obtained predictions for these perturbed instances are recorded.
4. **Interpretable model fitting**: Subsequently, LIME proceeds to fit an interpretable and transparent model, often a simple linear model, using the generated perturbed data and their corresponding model predictions. The aim is to construct a locally accurate model approximation that captures the decision-making process of the black-box model in the vicinity of the selected instance.

5. **Feature weighting**: The coefficients or weights derived from the interpretable model correspond to the influence or significance of each feature in the local decision boundary. These feature weights signify the contribution of individual features to the prediction for the selected instance.
6. **Explanation generation**: LIME finalizes the process by offering an explanatory account for the prediction of the chosen instance. This explanation is constructed by highlighting the feature weights associated with the interpretable model, indicating the relative importance of each feature in shaping the prediction outcome for that particular local context.

LIME's capacity for model-agnostic interpretability renders it adaptable to diverse ML and DL models without requiring access to their internal mechanisms. By establishing a simplified, interpretable model in proximity to a specific data point, LIME provides insights concerning the localized behavior of the complex model, facilitating human comprehension of why a particular prediction was rendered. This capability is particularly valuable in domains where transparency and accountability in decision-making are paramount, such as healthcare, finance, and legal domains.

The second most widely used framework for understanding the inner-workings of black-box models is SHAP, short for SHapley Additive exPlanations (Bertossi et al., 2021). SHAP proposed by Lundberg and Lee (2017) is based on cooperative game theory, and it presents a unified framework for interpreting predictions. Namely, the authors identified a new class of additive feature attribution methods, which are explanation models that have the form of a linear combination of binary variables. The paper shows that this class unifies six existing methods, such as LIME, DeepLIFT, and QII and it further shows that there is a unique solution in the class of additive feature attribution methods that satisfies four desirable properties of explanations: local accuracy, missingness, consistency, and linearity.

## 2.2. Utility of classical XAI methods for AI-based systems applied in finance

While SHAP and LIME have made substantial contributions to the field of XAI, they are not without their limitations. In this section, we will explore some of the key constraints and challenges associated with these classic XAI methods.

**Computational Complexity.** One of the most significant limitations of SHAP is its computational complexity. The calculation of SHAP values involves exploring all possible combinations of features, which grows exponentially with the number of features (Ribeiro et al. 2016, Bertossi et al., 2021). For complex models with numerous input features, the computational burden becomes prohibitively high. This limitation hinders the practical applicability of SHAP, especially in use cases containing high-dimensional data where real-time or near-real-time explanations are required. Similarly, LIME can also face computational challenges, albeit in a different context. Generating a perturbed dataset and training a local interpretable model for each prediction can be time-consuming, especially for large datasets and complex models, which are common-place in financial applications. This computational overhead can make LIME impractical for certain applications as well.

**Inherent model bias.** One of the fundamental challenges with SHAP and LIME is that they rely on surrogate models or approximations to explain complex black-box models. These surrogate models are typically simpler, interpretable models like linear regression or decision trees. However, these surrogate models may introduce their own biases and limitations, leading to inaccurate explanations (Gramegna and Giudici, 2021). For instance, if a black-box model exhibits non-linear behavior, attempting to explain it using a linear surrogate model can result in significant inaccuracies. Similarly, if the true relationship between features and predictions is highly complex, a simple surrogate model may fail to capture it adequately. This limitation underscores the trade-off between interpretability and accuracy in XAI methods like LIME.

**Sensitivity to data sampling.** In the context of LIME, this approach's effectiveness relies heavily on the quality and representativeness of the perturbed dataset used to train the local surrogate model. If the dataset is not sampled correctly or does not adequately cover the feature space, the explanations provided by LIME may be misleading or incomplete. Additionally, the choice of perturbation strategy and the number of perturbed samples can impact the stability and reliability of LIME explanations.

**Dealing with feature dependence.** Feature dependence means that the effect of one feature on the prediction depends on the values of other features. Both LIME and SHAP have limitations when dealing with feature dependence which in turn is a defining property of financial data. In the context of LIME, this method works by fitting a local linear model around the prediction to be explained, and using the coefficients of the linear model as feature importance scores. However, this approach assumes that the features are independent, which may not be true in reality. If the features are dependent, the linear model may not capture the true interactions and nonlinearities of the underlying model, and the feature importance scores may be inaccurate or misleading. On the other hand, as discussed previously, SHAP is based on the concept of Shapley values, which are derived from game theory and measure the marginal contribution of each feature to the prediction. Computing the exact Shapley values is computationally expensive, so SHAP uses various approximations to speed up the calculation. Specifically, conditional or interventional distributions are used for the purpose of computing the contribution of each feature to the output of the model (Kumar, 2020). The computation of both the conditional and interventional distributions assumes that the features are independent of each other. Hence, as argued by Molnar (2019), like many perturbation-based interpretation methods, the SHAP method suffers from inclusion of unrealistic data instances when features are correlated. We demonstrate the issue with a very simple example. Let's imagine that we train a model to predict apartment prices by considering many features among which the number of rooms and the apartments' size (in square feet) (Figure 1).

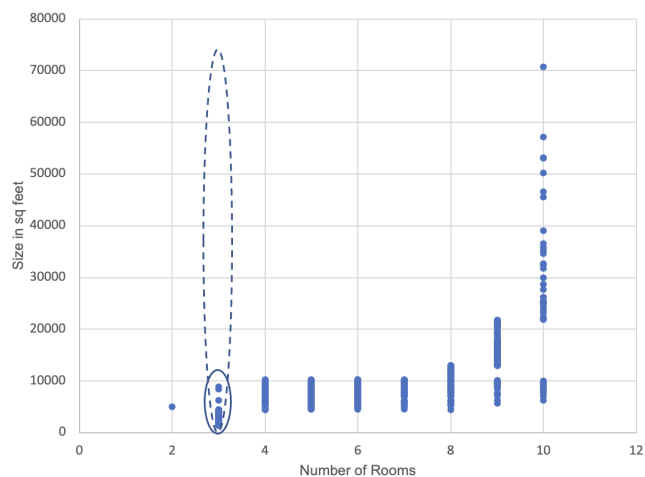

**Figure 1.** Issues with correlated features when calculating SHAP

Regardless of which underlining ML model we use, during the training phase, the model will only rely on realistic pairs. Yet, when calculating the SHAP values, in order to get the marginal contributions of each feature to the prediction, we will permute each feature across their entire range (this is represented by the dotted line in Figure 2) which leads to unrealistic pairing (i.e., a 70,000 sq feet apartment with 2.5 rooms). In observational studies and ML problems, it is very rare that the features are statistically independent (Aas et al., 2020). This is especially the case with financial data. If there is a high degree of dependence/correlation among some or all the features, the resulting explanations might be inaccurate hence research should focus on proposing new approaches that can provide meaningful explanations even in the case in which the input space contains features that are correlated.

**Stability of explanations provided.** Studies have repeatedly found that **explanations generated by attribution-based methods are not stable**, i.e., infinitesimal changes to the inputs can result in substantially different explanations (Dombrowski et al., 2019, Ghorbani et al., 2019; Slack et al., 2020; Bansal et al., 2020). In addition, there has been very little effort to ensure that the explanations produced by these methods are stable and robust. Notable exceptions include Alvarez-Melis and Jaakkola (2018), Fel et al. (2021) and Agarwal et al. (2022). Alvarez-Melis and Jaakkola (2018) propose a metric for stability based on the Lipschitz continuity, a parametric notion of stability that measures relative changes in the output with respect to the input. Fel et al. (2021) in turn propose two new measures for evaluating explanations borrowed from the field of algorithmic stability: mean generalizability MeGe and relative consistency ReCo. Finally, Agerwal et al. (2022) propose a relative measure that also leverages potentially meaningful information, such as the model's internal representation, to assess stability. Although these measures can be useful, they are rarely tested on real datasets, with none of them being tested on financial data. Moreover, the existing metrics are not enabled for an intuitive understanding of stability and robustness.

# 3. Good practices for AI-based Systems: What we know?

In the following section, we will outline a set of good practices for deploying explainability in AI systems within the finance sector. These practices aim to address the unique challenges and requirements of the financial industry, ensuring that AI-based models not only provide accurate predictions but also offer transparency, interpretability, and compliance with regulatory standards. By adhering to these practices, financial institutions can navigate the complex landscape of AI deployment in finance with confidence and effectiveness. Specifically, we explore the best practices focusing on data quality, audience-specific methods, and the consideration of data properties, particularly feature dependence.

**Data Quality is Crucial**. Data quality forms the bedrock of any AI-based system, and this holds especially true in the finance sector. Ensuring data is of high quality is the first and most essential step in deploying explainability effectively. Model developers must carefully select data sources and ensure data collection methods are robust and this involves validating data sources for accuracy, consistency, and completeness. Inaccurate or incomplete data can lead to misleading model outputs and explanations. Data cleaning and preprocessing are fundamental to ensure that the data used for training and inference is reliable. This includes handling missing values, outliers, and data inconsistencies. Finally, feature engineering plays a crucial role in shaping the input data. Financial experts should collaborate with data scientists to identify relevant features and create meaningful representations of financial data. Put differently, feature selection should be guided by domain knowledge to eliminate irrelevant or redundant variables.

**Match Explainability Methods with the Specific Audience**. Not all audiences have the same level of expertise or require the same level of detail in explanations. It is thus crucial to tailor explainability methods to the specific needs and expertise of the audience. For financial experts, who have a deep understanding of financial markets and instruments, explanations can be more technical and detailed. They may appreciate feature importance scores, partial dependence plots, and even access to model equations. Non-technical stakeholders, including customers, regulatory bodies, or C-suite executives, may require more intuitive and high-level explanations. Visualization tools, simple narratives, and summary statistics can be more effective in conveying insights to this audience. Regulators and auditors often require detailed documentation of model development, data sources, and validation procedures. They may also need access to specialized tools for auditing AI-based systems.

**Consideration of Data Properties and Feature Dependence.** Understanding the properties of the data being analyzed is critical when choosing explainability methods. Applying XAI methods without considerations of data characteristics like feature dependence can lead to misleading interpretations. In finance, features can be highly interdependent or multicollinear. Applying XAI methods designed for independent features to such data can produce ambiguous or contradictory explanations. Before applying XAI, it is essential to identify and address feature dependence through techniques like dimensionality reduction or feature selection. Furthermore, financial data is often time-series data, characterized by temporal dependencies and seasonality. Traditional XAI methods may not capture these dynamics effectively. Specialized time-series explainability techniques, such as the X-functions elaborated in Wildi and Hadji Misheva (2022) should be considered. Furthermore, financial markets are subject to changing conditions and trends. Models that work well under one set of conditions may not generalize to others. Explainability methods should consider the adaptability of models to non-stationary data and provide insights into when and why models perform differently under varying market conditions.

**Stability of Explanations.** A critical aspect of deploying explainability in AI systems for finance is ensuring the stability of explanations. XAI methods should be rigorously tested for their stability to ensure that the provided explanations remain consistent and reliable over time and across different datasets. In the dynamic environment of finance, where market conditions and data can change rapidly, the stability of explanations is paramount. By conducting thorough stability assessments, financial institutions can have greater confidence in the longevity and trustworthiness of their AI-based models, ultimately enhancing decision-making, risk assessment, and regulatory compliance. In this context, we argue that robustness to local perturbations of the input is a fundamental quality that explainability approaches should satisfy. In its most basic form, such a criterion means that similar units of analysis should not have explanations that are significantly different from one another (Alvarez-Melis and Jaakkola, 2018).

## 4. Conclusion Remarks

In the rapidly evolving landscape of financial prediction and forecasting using machine learning and deep learning, the deployment of explainability methods has become a crucial aspect. While the adoption of complex models offers significant advantages in terms of predictive accuracy, it also introduces challenges related to transparency and interpretability. This paper has highlighted the importance of good practices for deploying explainability in AI-based systems for finance, offering insights into how financial institutions can navigate these challenges effectively.

First and foremost, data quality is emphasized as the foundation of any AI-based system. Ensuring that data is accurate, consistent, and complete is paramount, as inaccurate or incomplete data can lead to misleading model outputs and explanations. Data preprocessing and feature engineering play essential roles in shaping the quality of input data. Tailoring explainability methods to the specific audience is another critical practice. Different stakeholders, including financial experts, non-technical audiences, regulators, and auditors, have varying levels of expertise and requirements for explanations. Providing explanations that are appropriate for the audience's needs enhances understanding and trust in AI-based systems. Consideration of data properties, particularly feature dependence, is highlighted as a crucial factor. Financial data often exhibits feature interdependence, multicollinearity, and time-series characteristics. Applying XAI methods without addressing these properties can result in inaccurate interpretations. Specialized techniques and approaches that account for these characteristics should be employed. Stability of explanations is recognized as a fundamental aspect of deploying explainability in finance. In the dynamic environment of financial markets, where conditions and data can change rapidly, ensuring the stability of explanations is crucial. XAI methods should be rigorously tested for stability, and robustness to local perturbations of the input should be a fundamental criterion.

In summary, by adhering to these good practices, financial institutions can effectively deploy explainability in AI-based systems, bridging the gap between complex models and actionable insights while ensuring compliance with regulatory standards and industry-specific requirements. The continued development and refinement of XAI methods tailored to the financial domain will play a pivotal role in enhancing transparency, accountability, and trust in AI-based decision-making processes within the financial sector.